# Competing striped structures in La$_2$CuO$_{4+y}$

Nicola Poccia[1], Alessandro Ricci[2], Gaetano Campi[3], A.-S. Caporale[4], Antonio Bianconi[4]

[1]*MESA, Institute for Nanotechnology University of Twente P.O. Box 217, 7500AE Enschede, Netherlands.*
[2]*Deutsches Elektronen-Synchrotron DESY, Notkestraße 85, D-22607 Hamburg, Germany*
[3]*Institute of Crystallography, CNR, Via Salaria Km 29.3, Monterotondo Stazione, Roma, I-00015 Italy.*
[4]*RICMAS, Rome International Center for Materials Science Superstripes Via dei Sabelli 119A, 00185 Roma, Italy.*

**Abstract**

High temperature superconductivity emerges in unique materials, like cuprates, that belong to the class of heterostructures at atomic limit, made of a superlattice of superconducting atomic layers intercalated by spacer layers. The physical properties of a strongly correlated electronic system, emerge from the competition between different phases with a resulting inhomogeneity from nanoscale to micron scale. Here we focus on the spatial arrangements of two types of structural defects in the cuprate La$_2$CuO$_{4+y}$ : i) the local lattice distortions in the CuO$_2$ active layers and ii) the lattice distortions around the charged chemical dopants in the spacer layers. We use a new advanced microscopy method: scanning nano X-ray diffraction (nXRD). We show here that local lattice distortions form incommensurate nanoscale ripples spatially anticorrelated with puddles of self-organized chemical dopants in the spacer layers.

**Introduction**

The science of granular materials, with typical particle size of few microns, has been interest of many outstanding scientists like Coulomb, Reynolds and de Gennes [1]. The interest is now focused on granular self organization of nanoscale puddles that determines unique functional properties of a complex material. Many works have been addressed to self organization of defects in multiferroics [2], graphene [3] and superconducting oxides [4-7] related with their magnetic or electronic functionality. In this interesting complex matter the nanoscale puddles are large enough to have a distinct electronic structure, but at the same time sufficiently small in order to exhibit quantum size effects [8,9]. The complex physics arise form self organization of nanoscale puddles over multiple scales from atomic-scale (0.1-1 nm) to nano-scale (1-100 nm) to meso-scale $10^2$-$10^3$ nm) to micro-scale (1-100 μm). This complex materials are in a metastable state therefore it is possible to tailor these granular heterogeneities with an external field to achieve better or new functionalities. This is becoming a field of research for its own as it has been shown recently to be feasible by using an intense X-ray beam in high-$T_c$ superconducting copper oxide materials [6,7]. Here X-ray illumination does not induce radiation damage and degradation, but ordering of oxygen interstitials with $T_c$ enhancements.

High temperature superconductivity emerges in a superlattice of atomic $CuO_2$ layers intercalated by spacer layers. The lattice of the atomic $CuO_2$ layers is inhomogeneous with lattice fluctuations away from the average structure. The local lattice distortions (LLD) at the atomic scale have been detected by fast and local experimental probes EXFAS, XANES and PDF [10-16]. Essential LLD are due to: i) pseudo-Jahn Teller polarons associated with doped holes in the $O(2p^5)Cu(3d^9)$ orbital, ii) the microstrain in $CuO_2$ layers lattice due to misfit strain between the active and spacer layers, and iii) the lattice distortions around the charged chemical dopants inserted in the spacer layers.

Local lattice distortions have been shown to increase below 200 K in $La_2CuO_{4+y}$ and have been correlated with the pseudo gap phase and polarons ordering [15]. These types of defects in a $La_2CuO_{4.06}$ single crystal self organize in spots left free by ordered oxygen interstitials puddles [17,18] like ripples in $Bi_2Sr_2CaCu_2O_{8+y}$ [19-23]. The defect organization works to establish the optimum inhomogeneity which raises the critical temperature to the optimum value [6,7,17,18]. The incommensurate short range superstructure satellites probing the local lattice distortions of the $CuO_2$ plane are similar to the satellites in the X-ray diffraction of Bi2212 [19-23] which have been shown to be due to the $CuO_2$ plane corrugation driven by the misfit strain among the layers [24-27] that induces a microstrain in the $CuO_2$ layers with a compression of the Cu-O bond length from the equilibrium Cu-O bond distance of 197 pm [28].

**Materials and methods**

Diffraction measurements on a $La_2CuO_{4.06}$ single crystal have been performed at the crystallography beamline XRD1 of ELETTRA third generation Synchrotron in Trieste (Italy). The sample was grown first as $La_2CuO_4$ by flux method and then doped by electrochemical oxidation. The X-ray beam, emitted by the wiggler source of 2 GeV electron storage ring, was monochromatized by a Si(111) double crystal, and focused on the sample. The temperature of the crystal was monitored with an accuracy of ±1K. We have collected the XRD data in the reflection geometry using a CCD area detector, a photon energy of 12.4 KeV (wavelength λ=1Å) and a high X-ray flux reaching a maximum value of $1.6*10^{14}$ photons/(sec cm$^2$). The sample oscillation around the b axis was in a range $0<\Phi<20°$, where Φ is the angle between the direction of the photon beam and the a axis. We have investigated a portion of the reciprocal space up to 0.6 Å$^{-1}$ momentum transfer, recording diffraction spots up to the maximum (3 3 19) reflex. Thanks to the high brilliance source, it has been possible to record a large number of weak superstructure spots around the main peaks of the average structure. Twinning of the crystal has been taken into account to index the superstructure peaks. The space group of the sample is Fmmm. Scanning nanoscale diffraction experiments were performed at the ID13 beamline of the European Synchrotron Radiation Facility (ESRF),

Grenoble, France. The ID13 nano-branch is specialized in the delivery of nano-focused X-ray beams for diffraction experiments. The photon source, a 18 mm period in-vacuum undulator works in the range 12-14 KeV with the storage ring operating at 6.03 GeV in the top-up mode with a current of 200 mA. The beamline is equipped with a Si-111 channel-cut crystal monochromator cooled with liquid nitrogen. A monochromatic X-ray beam of photon energy 14 keV was used, focused by Kirkpatrick Baez (KB) mirrors to a 300 nm spot at the sample (full width at half maximum). A 16 bit two-dimensional Fast Readout Low Noise charged coupled device (FReLoN CCD) detector with 2048 x 2048 pixels of 51 x 51 $\mu m^2$ was used, binned to 512 x 512 pixels. The detector was placed 60 mm behind the sample and offset. Diffraction images were obtained after correcting the 2D images for dark noise, flat field, distortion and eventually background. The CCD camera records the intensity of the satellite superstructures. The intensity is integrated over square subareas of the images recorded by the CCD detector in reciprocal-lattice units (r.l.u.) and then normalized to the intensity ($I_0$) of the tail of the main crystalline reflections at each point (x,y) of the sample reached by the translator. The interaction volume of the 300 nm impinging beam with the crystal is 0.3 x 0.3 x 1.5 $\mu m^3$.

This technique gives a mixed information of the k-space and r-space of the bulk structure heterogeneities and only recently it has been applied to cuprate oxide superconductors. In particular, it allows to investigate spatial heterogeneity of ordered puddles combining high wave number resolution with micrometer spatial resolution in a similar way as the transmission electron microscopy (TEM), but without the complication of electron beam damage. In fact, the heating of the thin crystals caused by the electron beam may change the mobile oxygen content y and generate transient non-equilibrium surface structures masking intrinsic bulk effects. Therefore quantitative details about the finite size ordering properties and the temperature dependent studies are not possible by TEM.

**Results and Discussion**

Using scanning nXRD we have been able to probe the spatial distribution of the two superstructure satellites $q_2$ = 0.25 **b***+0.5 **c*** and $q_3$ = 0.21 **b***+0.29 **c***. Figure 1 (panel a) shows a schematic view of the ordering of oxygen interstitials forming wires in the a crystallographic direction with $q_2$ wave-vector. These wires have a periodicity of 4 lattice units in the b axis and 2 lattice units in the c* axis. The droplets determined by the self organization of oxygen interstitials (O$i$) with superlattice wave-vector $q_2$, shown in Figure 1 a, are named with the acronym Q2-O$i$. Figure 1 (panel b) shows a schematic view of the ordering of local lattice distortions forming wires along the a* crystallographic direction. These wires have a periodicity of 4.7 lattice units in the b* axis and 3.4 lattice units in the c* axis directions. The incommensurate $q_3$ reflections are weak and diffuse indicating that the size of the local lattice distortion droplets is of the order of 20 nm. The droplets determined by the self organization of local lattice distortions (LLD) with superlattice wave-vector $q_3$, shown in Figure 2 b, are named with the acronym Q3-LLD.

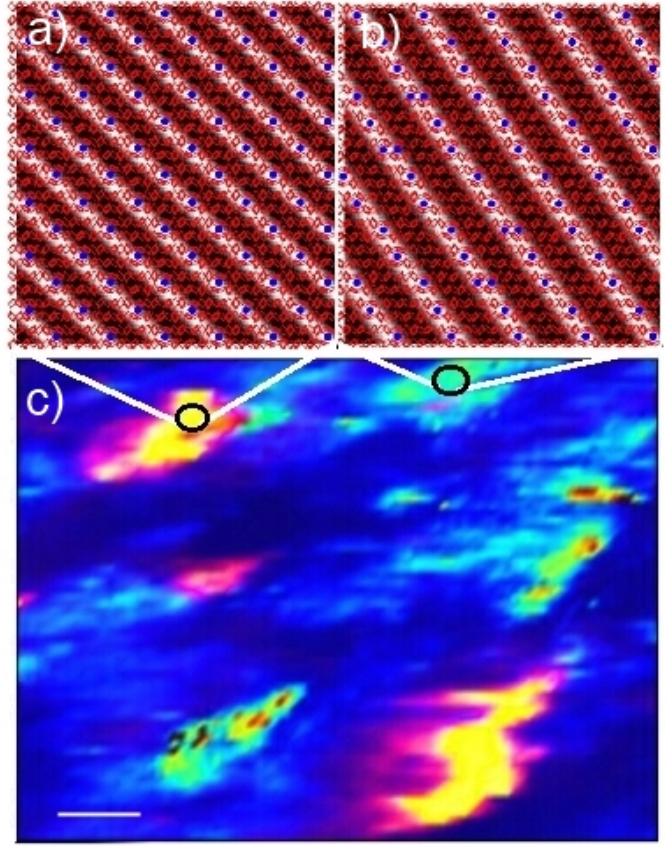

**Figure 1.** Panel a) schematic view of the long range superstructure satellites $q_2 = 0.25\ b^* + 0.5\ c^*$, probing the Q2-Oi puddles, made of ordered oxygen interstitials. Panel b) Q3-LLD puddles of local lattice distortions, probed by the X-ray superstructure $q_3 = 0.21\ b^* + 0.29\ c^*$ (black filled circles). The horizontal and the vertical directions correspond respectively to the b and c crystallographic axis. Panel c) real space position dependence of the Q2-Oi (hot colors) and the Q3-LLD (cold colors) puddles. The horizontal and the vertical directions correspond respectively to the a and b crystallographic axis. The white bar represents a distance of 20 nm. The Q2-Oi and Q3-LLD puddles occupy different portions of the real space.

The distribution of the Q2-O$i$ and Q3-LLD puddles in real space are shown in Fig. 1 (panel c) where the imaging of the regions containing incommensurate modulated local lattice distortions (LLD) and oxygen interstitials (O$i$) are represented in a 2D colour plot. The cold colours correspond to the areas occupied by ordered local lattice distortions while the hot ones indicate regions rich of ordered oxygen interstitials. The dark blue areas of the mapping do not show either ordered domains of LLD or O$i$. Although we see regions of the sample with an high intensity of the $q_3$ reflections on the micron-scale, the size of each Q3-LLD domain remains of the order of 20 nm. Therefore the strongly diffracting areas indicate regions with a high density of Q3-LLD droplets. On the contrary, the Q2-O$i$ domains, grow in size as a function of the intensity of the peak, and the full width half maximum of the $q_2$ superstructure. The oxygen interstitials form a granular system of superconducting grains percolating at optimum doping [7,17,18].

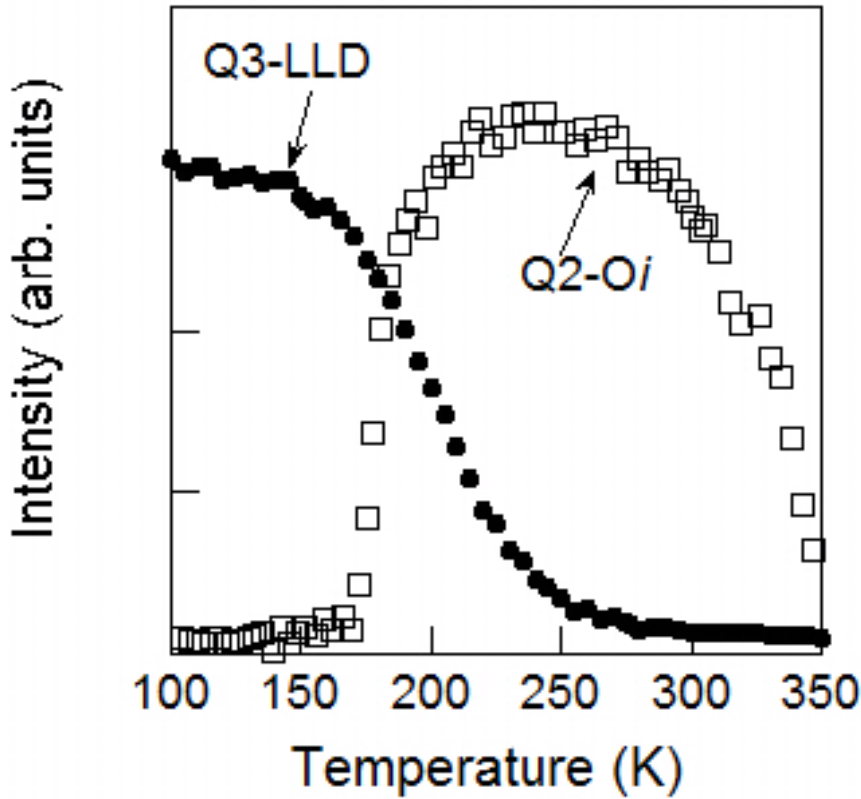

**Figure 2.** Thermal treatment of a quenched $La_2CuO_{4+y}$ single crystal, starting at 100 K with a fully disordered oxygen interstitial phase. The black filled circles shows the evolution of the intensity of the superstructure satellites $q_3 = 0.21 \mathbf{b^*}+0.29 \mathbf{c^*}$ as a function of temperature probing the density of nanoscale Q3-LLD puddles of local lattice distortions. The empty black squares show the intensity of the long range superstructure satellites $q_2 = 0.25 \mathbf{b^*}+0.5 \mathbf{c^*}$ probing the puddles of Q2-Oi oxygen interstitial.

Fig. 2 shows the response of the Q3-LLD droplets and Q2-O$i$ puddles to thermal treatments. We have obtained a sample with only Q3-LLD droplets by increasing the temperature to above 350 K for which oxygen interstitials get disordered. This can be done through a rapid quench of the sample below 200 K in such a way that the oxygen interstitial remain disordered at low temperature. Figure 2 shows the thermal evolution of the oxygen interstitials, measured by X-ray diffraction performed on the crystallography beamline, XRD1, at ELETTRA. The sample temperature has been slowly increased from 100 K to 350 K. The evolution of the Q3-LLD droplets and Q2-O$i$ puddles as a function of temperature supports the different origin for these two kind of defects. The Q2-O$i$ puddles start to form at about 180 K since warming the sample, the oxygen interstitials that originally were in a glassy phase, get mobile. The maximum ordering is reached at 250 K and for higher temperature it decreases. This indicates the melting of the Q2-O$i$ puddles. The Q2-Oi puddles order show an hysteresis in the order–disorder transition controlled by irradiating the sample by a continuous X-ray flux [29]. The oxygen interstitial grains grow in an anisotropic way under X-ray illumination, first in the a–b plane and later in the c-axis direction [6,29]. On the contrary the diffraction signal, from the competing Q3-LLD droplets, shows a continuous decreasing from 180 K to 250 K followed by a slowly decreasing signal at higher temperatures.

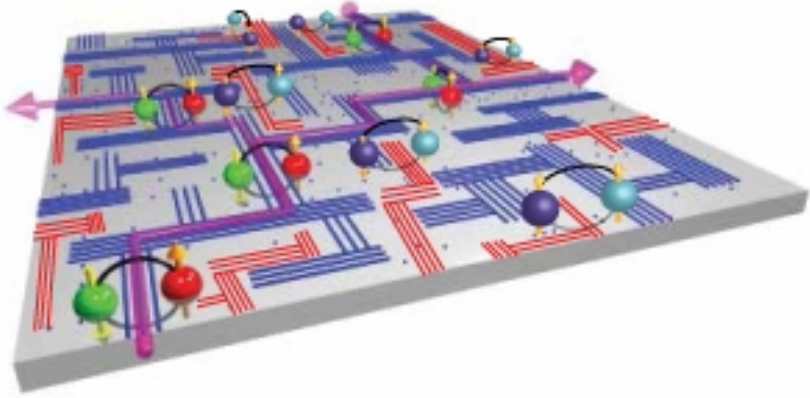

**Figure 3.** Pictorial view of the two networks of defects made respectively of local lattice distortions (red stripes) and oxygen interstitials (blue stripes). The nanoscale stripes in a optimum doped $La_2CuO_{4+y}$ single crystal form two percolating networks which stabilize the optimum superconducting critical temperature. Distinct Cooper-pairs, on the underling nanoscale striped lattice, are shown to travel along different pathways.

The high $T_c$ superconducting phase is established clearly at low temperature in a very complex granular matter. Figure 3 shows a pictorial scheme of the complex pattern of the Q3-LLD striped droplets and Q2-O$i$ striped puddles which appears to occur in $La_2CuO_{4+y}$ at all doping regimes. The intertwining of two structural units : striped oxygen interstitials pattern (blue stripes) and striped local distortions pattern (red-stripes) form a percolative network. When both patterns exhibit a power-law distribution of the domains sizes [6, 7, 17, 18] the system shows the maximum critical temperature. What we call optimum inhomogeneity [7] is therefore the spatial statistical arrangement of defects which follow a perfect power-law distribution. These results demonstrate that heterogeneous granular phases occur in a typical simple cuprate like in iron based superconductors [30-39]. The networks of nanosized grains are expected to give a superconducting phase controlled by Josephson-like links as in the arrays of superconducting dots [40]. Our experimental results prove that the lattice of HTS has a nanoscale complex granular structure. These data support theories having taken into account the intrinsic granularity in high temperature superconductors [41-56] and the multi-band, multi-gap or multi-condensate superconductivity [8,9,57-65]. Although in solid-state physics multi-component nanoscale granular matter for novel material functionalities has only recently started to be explored, in biology multi-scale structural organization is already known to be essential in the organization of the living matter [66-67].

**Conclusions**

From a conceptual point of view, an attractive feature of granular materials such as $La_2CuO_{4+y}$ single crystal is the possibility to separately control the effects of electron interaction and quantum interference. These results give an unprecedented outlook on how the self-organize local lattice distortions 'Q3' and oxygen interstitials 'Q2' sets the quantum interferences required to establish high temperature superconductivity. The observation of these complex patterns in high temperature superconductors could be an interesting starting point for the developing of new nano-devices.